\documentstyle[12pt]{article}

\begin{document}

\author{C. Bizdadea and S. O. Saliu\thanks{%
e-mail addresses: odile.saliu@comp-craiova.ro or osaliu@central.ucv.ro} \\
Department of Physics, University of Craiova\\
13 A. I. Cuza Str., Craiova R-1100, Romania}
\title{Irreducible Antifield Analysis of $p$-Form Gauge Theories }
\maketitle

\begin{abstract}
The irreducible antifield formalism for $p$-form gauge theories with gauge
invariant interaction terms is exposed. The ghosts of ghosts do not appear.
The acyclicity of the Koszul-Tate operator is ensured without introducing
antifields at resolution degrees higher that two.

\end{abstract}

It is widely known that gauge theories involving $p$-forms are important due
to their link with string theory and supergravity models \cite{1}--\cite{6},
and also due to the importance of their characteristic cohomology \cite{7}.
A typical feature of such theories is given by their redundant behavior,
i.e., the reducibility of the gauge generators. The reducibility further
implies the presence of ghost fields with ghost number greater that one
(ghosts of ghosts, etc.), and, in the meantime, a pyramid of non-minimal
variables in the framework of the Lagrangian BRST formalism \cite{8}--\cite
{12}. The ghost and non-minimal structures of interacting $p$-forms were
exposed in \cite{17} in the light of the reducible Hamiltonian BRST method.

This paper investigates whether gauge theories involving abelian $p$-forms
with interactions that possess the same gauge invariances like the free
theory can be consistently quantized along the irreducible antifield-BRST
formalism. Our main result is that this is always possible (for the priorly
invoked class of theories). As far as we know, this problem has not been
approached. As a consequence of our irreducible treatment, the ghosts of
ghosts are not necessary, and the auxiliary fields have a linear structure
(in contrast with the pyramidal reducible structure). Our method basically
relies on replacing the redundant gauge theory with an irreducible one
possessing the same physical observables, and on further quantizing the
resulting irreducible gauge system.

The main points approached in our paper are as follows. First, we explain in
detail our irreducible mechanism in the case of free abelian $p$-form gauge
fields. Second, we extend the results to gauge theories involving various
sorts of abelian forms with gauge invariant interaction terms, and reduce
the problem to the free case analysis. This is possible by virtue of the
fact that the reducibility functions associated with the interacting system
are diagonal.

We start with the Lagrangian action of free abelian $p$-form gauge fields ($%
p>1$) 
\begin{equation}
\label{1}S_{0_p}^L\left[ A_{(p)}^{\mu _1\ldots \mu _p}\right] =-\frac
1{2\cdot \left( p+1\right) !}\int d^DxF_{(p)\mu _1\ldots \mu
_{p+1}}F_{(p)}^{\mu _1\ldots \mu _{p+1}}, 
\end{equation}
where $F_{(p)\mu _1\ldots \mu _{p+1}}$ are the field strengths of the
antisymmetric fields $A_{(p)\mu _1\ldots \mu _p}$ and $D>\left( p+1\right) $%
. The additional index $(p)$ is introduced by virtue of the subsequent
discussion, where various sorts of abelian $p$-form gauge fields will be
dealt with. Action (\ref{1}) is invariant under the gauge transformations 
\begin{equation}
\label{2}\delta _\epsilon A_{(p)}^{\mu _1\ldots \mu _p}=\partial _{}^{\left[
\mu _1\right. }\epsilon _{(p)}^{\left. \mu _2\ldots \mu _p\right]
}=Z_{\;\;\nu _1\ldots \nu _{p-1}}^{\mu _1\ldots \mu _p}\epsilon _{(p)}^{\nu
_1\ldots \nu _{p-1}}, 
\end{equation}
with $\left[ \mu _1\ldots \mu _p\right] $ denoting antisymmetry with respect
to the indices between brackets. This model is $\left( p-1\right) $-stage
reducible, the reducibility relations 
\begin{equation}
\label{3}Z_{\;\;\nu _1\ldots \nu _{p-k-1}}^{\mu _1\ldots \mu
_{p-k}}Z_{\;\;\lambda _1\ldots \lambda _{p-k-2}}^{\nu _1\ldots \nu
_{p-k-1}}=0, 
\end{equation}
holding off-shell, where the $k$th order reducibility functions are given by 
\begin{equation}
\label{4}Z_{\;\;\nu _1\ldots \nu _{p-k-1}}^{\mu _1\ldots \mu _{p-k}}=\frac
1{\left( p-k-1\right) !}\partial _{}^{\left[ \mu _1\right. }\delta _{\;\;\nu
_1}^{\mu _2}\ldots \delta _{\;\;\nu _{p-k-1}}^{\left. \mu _{p-k}\right]
},\;k=0,\ldots ,p-1. 
\end{equation}

First, we introduce a bosonic antisymmetric field, $A_{(p)}^{\lambda
_1\ldots \lambda _{p-k-2}}$, associated with every reducibility relation (%
\ref{3}) corresponding to $k\geq 0$ even, and also a bosonic antisymmetric
gauge parameter, $\epsilon _{(p)}^{\lambda _1\ldots \lambda _{p-k-2}}$,
attached to every reducibility relation (\ref{3}) corresponding to $k\geq 1$
odd. We require the next gauge transformations for the new fields 
\begin{equation}
\label{51}\delta _\epsilon A_{(p)}^{\mu _1\ldots \mu _{p-2k}}=\partial
_{}^{\left[ \mu _1\right. }\epsilon _{(p)}^{\left. \mu _2\ldots \mu
_{p-2k}\right] }+\left( p-2k+1\right) \partial _\mu \epsilon _{(p)}^{\mu \mu
_1\ldots \mu _{p-2k}},\;k=1,\ldots ,b_p, 
\end{equation}
with $b_p=p/2$, or $\left( p-1\right) /2$ for $p$ even, respectively, odd.
Here and throughout this paper, we use the conventions $f^{\mu _1\ldots \mu
_m}=f$ if $m=0$, and $f^{\mu _1\ldots \mu _m}=0$ if $m<0$. We consider the
theory described by the Lagrangian action 
\begin{equation}
\label{52}S_{0_p}^L\left[ A_{(p)}^{\mu _1\ldots \mu _p},A_{(p)}^{\mu
_1\ldots \mu _{p-2k}}\right] =S_{0_p}^L\left[ A_{(p)}^{\mu _1\ldots \mu
_p}\right] , 
\end{equation}
subject to the gauge transformations (\ref{2}) and (\ref{51}). It is obvious
that (\ref{52}) is invariant under (\ref{51}).

Let us show that the above mentioned gauge transformations are irreducible.
In this respect, we take $\epsilon _{(p)}^{\mu _1\ldots \mu
_{p-2k-1}}=\partial _{}^{\left[ \mu _1\right. }\theta _{(p)}^{\left. \mu
_2\ldots \mu _{p-2k-1}\right] }$ in (\ref{2}) and (\ref{51}) (for $%
k=0,\ldots ,a_p$, with $a_p=p/2-1$, or $\left( p-1\right) /2$ for $p$ even,
respectively, odd), and find $\delta _\epsilon A_{(p)}^{\mu _1\ldots \mu
_p}=0$ and $\delta _\epsilon A_{(p)}^{\mu _1\ldots \mu _{p-2k}}=\left(
p-2k+1\right) \partial _\nu \partial _{}^{\left[ \nu \right. }\theta
_{(p)}^{\left. \mu _1\ldots \mu _{p-2k}\right] }$, $k>0$. The prior gauge
transformations vanish if and only if $\theta _{(p)}^{\mu _1\ldots \mu
_{p-2k}}=\partial _{}^{\left[ \mu _1\right. }\xi _{(p)}^{\left. \mu _2\ldots
\mu _{p-2k}\right] }$, so $\epsilon _{(p)}^{\mu _1\ldots \mu _{p-2k-1}}=0$.
In conclusion, $\delta _\epsilon A_{(p)}^{\mu _1\ldots \mu _{p-2k}}=0$ if
and only if $\epsilon _{(p)}^{\mu _1\ldots \mu _{p-2k-1}}=0$. This clearly
emphasizes that (\ref{2}) and (\ref{51}) are irreducible. In this way, we
associated an irreducible theory, described by action (\ref{52}) and the
gauge transformations (\ref{2}), (\ref{51}), with the starting redundant
model.

Next, we prove that both the irreducible and initial reducible theories
display the same physical observables. Let $F\left( A_{(p)}^{\mu _1\ldots
\mu _p},A_{(p)}^{\mu _1\ldots \mu _{p-2k}}\right) $ be an observable for the
irreducible system. As any physical observable must be gauge invariant, on
behalf of (\ref{2}) and (\ref{51}) it follows 
\begin{equation}
\label{57}\partial _{_{\left[ \mu _1\right. }}\frac{\delta F}{\delta
A_{(p)}^{\left. \mu _2\ldots \mu _{p-2k+1}\right] }}+\left( p-2k+2\right)
\partial ^\nu \frac{\delta F}{\delta A_{(p)}^{\nu \mu _1\ldots \mu _{p-2k+1}}%
}=0, 
\end{equation}
with $k=0,\ldots ,b_p$. We solve the above system starting from the last
equation. We explain the case $p$ even, the other situation being similar.
For $k=p/2$, equation (\ref{57}) reads $\partial _\mu \frac{\delta F}{\delta
A_{(p)}}+2\partial ^\nu \frac{\delta F}{\delta A_{(p)}^{\nu \mu }}=0$, which
implies $\partial ^\mu \partial _\mu \frac{\delta F}{\delta A_{(p)}}=0$.
Because $\partial ^\mu \partial _\mu $ is invertible, we obtain $\frac{%
\delta F}{\delta A_{(p)}}=0$, hence $\partial ^\nu \frac{\delta F}{\delta
A_{(p)}^{\nu \mu }}=0$. The next equation from (\ref{57}), $\partial
_{_{\left[ \mu _1\right. }}\frac{\delta F}{\delta A_{(p)}^{\left. \mu _2\mu
_3\right] }}+4\partial ^\nu \frac{\delta F}{\delta A_{(p)}^{\nu \mu _1\mu
_2\mu _3}}=0$, leads to $\partial ^{\mu _1}\partial _{_{\left[ \mu _1\right.
}}\frac{\delta F}{\delta A_{(p)}^{\left. \mu _2\mu _3\right] }}=0$, which
further implies, using $\partial ^\nu \frac{\delta F}{\delta A_{(p)}^{\nu
\mu }}=0$, that $\partial ^\mu \partial _\mu \frac{\delta F}{\delta
A_{(p)}^{\mu _1\mu _2}}=0$, so $\frac{\delta F}{\delta A_{(p)}^{\mu _1\mu _2}%
}=0$. Step by step, we infer along the same line that $F$ does not depend on
the new fields, $\frac{\delta F}{\delta A_{(p)}^{\mu _1\ldots \mu _{p-2k}}}%
=0 $ for $k=1,\ldots ,p/2$, such that the first equation from (\ref{57})
becomes $\partial _{_{\left[ \mu _1\right. }}\frac{\delta F}{\delta
A_{(p)}^{\left. \mu _2\ldots \mu _{p+1}\right] }}=0$. Due to the fact that,
on the one hand the last equations stand for the equation fulfilled by the
observables of the reducible theory, and on the other hand $F$ depends only
on the original fields, it results that any observable of the irreducible
system is also an observable of the reducible one. At the same time, it is
obvious that if $\bar F\left( A_{(p)}^{\mu _1\ldots \mu _p}\right) $
represents an observable of the reducible theory, then it remains so for the
irreducible system, because it automatically satisfies (\ref{57}). In
consequence, the irreducible and reducible versions are physically
equivalent, describing the same physical theory. From the point of view of
the BRST formalism, the zeroth order cohomological groups of the
longitudinal exterior derivative along the gauge orbits associated with the
reducible, respectively, irreducible model coincide. Thus, if we show that
the Koszul-Tate operator in the irreducible case is acyclic, the homological
perturbation theory \cite{13}--\cite{16} ensures: (i) the nilpotency of the
irreducible BRST symmetry, $s$, and (ii) $H^0\left( s\right) =\left\{ {\rm %
observables}\right\} $, with $H^0\left( s\right) $ the zeroth order
cohomological group of $s$. This makes legitimate from the physical point of
view, i.e., from the point of view of the requirements $s^2=0$, $H^0\left(
s\right) =\left\{ {\rm observables}\right\} $, the replacement of the BRST
quantization of the reducible theory with the quantization of the
irreducible one.

At this point we briefly investigate the higher order cohomological groups
of the longitudinal exterior derivative along the gauge orbits. This
analysis will be relevant during the gauge-fixing process. More precisely,
we show that all the higher order groups in the irreducible situation are
trivial. This can be seen by introducing the minimal ghost spectrum in the
irreducible case, namely $\eta _{(p)}^{\mu _1\ldots \mu _{p-2k-1}}$ (with $%
k=0,\ldots ,a_p$), and defining the action of the longitudinal exterior
derivative along the gauge orbits as usually, through 
\begin{equation}
\label{58b}DA_{(p)}^{\mu _1\ldots \mu _p}=\partial ^{\left[ \mu _1\right.
}\eta _{(p)}^{\left. \mu _2\ldots \mu _p\right] }, 
\end{equation}
\begin{equation}
\label{58c}DA_{(p)}^{\mu _1\ldots \mu _{p-2k}}=\partial ^{\left[ \mu
_1\right. }\eta _{(p)}^{\left. \mu _2\ldots \mu _{p-2k}\right] }+\left(
p-2k+1\right) \partial _\nu \eta _{(p)}^{\nu \mu _1\ldots \mu
_{p-2k}},\;k=1,\ldots ,b_p, 
\end{equation}
\begin{equation}
\label{58d}D\eta _{(p)}^{\mu _1\ldots \mu _{p-2k-1}}=0,\;k=0,\ldots ,a_p. 
\end{equation}
The ghosts $\eta $ are fermionic, with pure ghost number one. From (\ref{58b}%
--\ref{58c}) it will follow that all the ghosts are $D$-exact. Here we
indicate the line corresponding to $p$ even, the opposite situation being
treated similarly. We start with the last two equations from (\ref{58c}), $%
DA_{(p)}=\partial _\mu \eta _{(p)}^\mu $, $DA_{(p)}^{\mu \nu }=\partial
^{\left[ \mu \right. }\eta _{(p)}^{\left. \nu \right] }+3\partial _\rho \eta
_{(p)}^{\rho \mu \nu }$. Applying $\partial _\mu $ on the last relation and
using the first one, we derive $\eta _{(p)}^\mu =D\left( \frac 1{\Box
}\left( \partial _\nu A^{\nu \mu }+\partial ^\mu A\right) \right) $, such
that $\partial _\rho \eta _{(p)}^{\rho \mu \nu }=\frac 13D\left( \frac
1{\Box }\partial _\rho \partial ^{\left[ \rho \right. }A_{(p)}^{\left. \mu
\nu \right] }\right) $. Multiplying the next equation from (\ref{58c}) by $%
\partial _\mu $ and taking into account the expression of $\partial _\rho
\eta _{(p)}^{\rho \mu \nu }$, we obtain $\eta _{(p)}^{\mu \lambda \rho
}=D\left( \frac 1{\Box }\left( \partial _\nu A^{\nu \mu \lambda \rho }+\frac
13\partial ^{\left[ \mu \right. }A^{\left. \lambda \rho \right] }\right)
\right) $. Acting like before, we infer 
\begin{equation}
\label{58e}\eta _{(p)}^{\mu _1\ldots \mu _{p-2k-1}}=D\left( \frac 1{\Box
}\chi _{(p)}^{\mu _1\ldots \mu _{p-2k-1}}\right) ,\;k=0,\ldots ,a_p, 
\end{equation}
with $\chi _{(p)}^{\mu _1\ldots \mu _{p-2k-1}}=\partial _\mu A_{(p)}^{\mu
\mu _1\ldots \mu _{p-2k-1}}+\frac 1{p-2k-1}\partial ^{\left[ \mu _1\right.
}A_{(p)}^{\left. \mu _2\ldots \mu _{p-2k-1}\right] }$. On the one hand, from
(\ref{58d}) it follows that any $D$-closed quantity of pure ghost number
greater than zero is a polynomial in the ghosts with coefficients that are
gauge invariant functions. On the other hand, from (\ref{58e}) it results
that any such polynomial is $D$-exact. As a consequence, all $H^q\left(
D\right) $ with $q>0$ vanish in our irreducible approach.

Now, we prove that the Koszul-Tate operator corresponding to the irreducible
model, $\delta $, is truly acyclic. Accordingly the standard BRST receipt,
the minimal antifield spectrum reads as $A_{(p)\mu _1\ldots \mu _{p-2k}}^{*}$
($k=0,\ldots ,b_p$) and $\eta _{(p)\mu _1\ldots \mu _{p-2k-1}}^{*}$ (with $%
k=0,\ldots ,a_p$). The $A^{*}$'s are fermionic and have antighost number
one, while the $\eta ^{*}$'s are bosonic, with antighost number two. We
define the action of $\delta $ on the generators from the BRST complex as
usually by 
\begin{equation}
\label{60a}\delta A_{(p)}^{\mu _1\ldots \mu _{p-2k}}=0,\;k=0,\ldots
,b_p,\;\delta \eta _{(p)}^{\mu _1\ldots \mu _{p-2k-1}}=0,\;k=0,\ldots ,a_p, 
\end{equation}
\begin{equation}
\label{60b}\delta A_{(p)\mu _1\ldots \mu _p}^{*}=-\frac 1{p!}\partial ^\nu
F_{(p)\nu \mu _1\ldots \mu _p}, 
\end{equation}
\begin{equation}
\label{60c}\delta A_{(p)\mu _1\ldots \mu _{p-2k}}^{*}=0,\;k=1,\ldots ,b_p, 
\end{equation}
\begin{equation}
\label{60d}\delta \eta _{(p)\mu _1\ldots \mu _{p-2k-1}}^{*}=-\left( \left(
p-2k\right) \partial ^\mu A_{(p)\mu \mu _1\ldots \mu _{p-2k-1}}^{*}+\partial
_{\left[ \mu _1\right. }A_{(p)\left. \mu _2\ldots \mu _{p-2k-1}\right]
}^{*}\right) . 
\end{equation}
From (\ref{60b}) and (\ref{60c}), we observe that $\partial ^{\mu
_1}A_{(p)\mu _1\ldots \mu _p}^{*}$ and $A_{(p)\mu _1\ldots \mu _{p-2k}}^{*}$
are $\delta $-closed quantities. This implies that there can exist $\delta $%
-closed polynomials in the above objects with coefficients that may involve
the $A_{(p)}^{\mu _1\ldots \mu _{p-2k}}$'s in the higher order resolution
degree homological groups of $\delta $. In the reducible BRST approach, the $%
\delta $-closed polynomials are killed step by step in the homology of $%
\delta $ by appropriately introducing some new generators (the antifields
associated with the ghosts of ghosts). This is not necessary in our
treatment, as we show that the $\delta $-closed quantities with positive
resolution degrees from the irreducible case are $\delta $-exact. In this
respect, we prove that $\partial ^{\mu _1}A_{(p)\mu _1\ldots \mu _p}^{*}$
and $A_{(p)\mu _1\ldots \mu _{p-2k}}^{*}$ are $\delta $-exact. The proof
will be exemplified in the case $p$ even, being similar if $p$ odd, and goes
as follows. We start with the last equation from (\ref{60d}), $\delta \eta
_{(p)\mu _1}^{*}=-\left( 2\partial ^\mu A_{(p)\mu \mu _1}^{*}+\partial _{\mu
_1}A_{(p)}^{*}\right) $, that leads to $\delta \left( \partial ^\mu \eta
_{(p)\mu }^{*}\right) =-\partial ^\mu \partial _\mu A_{(p)}^{*}$, so $%
A_{(p)}^{*}=\delta \left( \frac{-1}{\Box }\partial ^\mu \eta _{(p)\mu
}^{*}\right) $. Thus, we have that $\partial ^\mu A_{(p)\mu \mu
_1}^{*}=\frac 12\delta \left( \frac 1{\Box }\partial _{\mu _1}\partial ^\mu
\eta _{(p)\mu }^{*}-\eta _{(p)\mu _1}^{*}\right) $. Using the last relation
in the next equation from (\ref{60d}) multiplied by $\partial ^{\mu _1}$, we
find $A_{(p)\mu _1\mu _2}^{*}=\delta \left( \frac{-1}{\Box }\left( \frac
12\partial _{\left[ \mu _1\right. }\eta _{(p)\left. \mu _2\right]
}^{*}+\partial ^\mu \eta _{(p)\mu \mu _1\mu _2}^{*}\right) \right) $. Along
the same line, we derive 
\begin{equation}
\label{60e}A_{(p)\mu _1\ldots \mu _{p-2k}}^{*}=\delta \left( \frac{-1}{\Box }%
\left( \frac 1{p-2k}\partial _{\left[ \mu _1\right. }\eta _{(p)\left. \mu
_2\ldots \mu _{p-2k}\right] }^{*}+\partial ^\mu \eta _{(p)\mu \mu _1\ldots
\mu _{p-2k}}^{*}\right) \right) , 
\end{equation}
for $k=1,\ldots ,\frac p2$, and 
\begin{equation}
\label{60f}\partial ^\mu A_{(p)\mu \mu _1\ldots \mu _{p-1}}^{*}=\delta
\left( \frac{-1}{p\,\Box }\left( \partial ^\mu \partial _{\left[ \mu \right.
}\eta _{(p)\left. \mu _1\ldots \mu _{p-1}\right] }^{*}\right) \right) . 
\end{equation}
Formulas (\ref{60e}--\ref{60f}) restore the $\delta $-exactness of the
investigated quantities. Moreover, we cannot find $\delta $-closed
expressions involving $\eta _{(p)\mu _1\ldots \mu _{p-2k-1}}^{*}$ due to the
irreducibility of the gauge transformations (\ref{2}) and (\ref{51}). In
conclusion, the above antifield spectrum is enough to enforce the acyclicity
of $\delta $, so we do not need to introduce antifields with resolution
degrees higher that two. In this way, the acyclicity of $\delta $ is fully
guaranteed within our irreducible approach, hence, as explained above, we
can replace the quantization of the reducible model with the one of the
irreducible theory.

In the sequel we perform the antifield-BRST quantization of the irreducible
system built previously. With the minimal ghost and antifield spectra at
hand, we choose the non-minimal sector $\left( \bar \eta _{(p)}^{\mu
_1\ldots \mu _{p-2k-1}},\bar \eta _{(p)\mu _1\ldots \mu
_{p-2k-1}}^{*}\right) $, $\left( B_{(p)}^{\mu _1\ldots \mu
_{p-2k-1}},B_{(p)\mu _1\ldots \mu _{p-2k-1}}^{*}\right) $ with $k=0,\ldots
,a_p$. The $B$'s and $\bar \eta ^{*}$'s are bosonic and with ghost number
zero, while the remaining variables are fermionic, of ghost number minus
one. The ghost number is defined as the difference between the pure ghost
number and the antighost number. The non-minimal solution of the master
equation is expressed by 
\begin{eqnarray}\label{63}
& &S^{(p)}=S_{0_p}^L+\int d^Dx\left( A_{(p)\mu _1\ldots \mu _p}^
{*}\partial ^{\left[ \mu _1\right. }\eta _{(p)}^{\left. \mu _2\ldots 
\mu _p\right] }+\sum_{k=0}^{a_p}\bar \eta _{(p)\mu _1\ldots 
\mu _{p-2k-1}}^{*}B_{(p)}^{\mu _1\ldots \mu _{p-2k-1}}\right. \nonumber \\
& &\left. +\sum_{k=1}^{b_p}A_{(p)\mu _1\ldots \mu _{p-2k}}^{*}
\left( \partial ^{\left[ \mu _1\right. }\eta _{(p)}^{\left. 
\mu _2\ldots \mu _{p-2k}\right] }+\left( p-2k+1\right) 
\partial _\mu \eta _{(p)}^{\mu \mu _1\ldots \mu _{p-2k}}\right) \right) . 
\end{eqnarray}  In order to fix the gauge, we recall formula (\ref{58e}),
which actually shows how the ghosts remove the unphysical degrees of
freedom. Indeed, we can regard the functions $\chi _{(p)}^{\mu _1\ldots \mu
_{p-2k-1}}$ like some fields with the gauge transformations $\delta _{\bar
\epsilon }\chi _{(p)}^{\mu _1\ldots \mu _{p-2k-1}}=\bar \epsilon _{(p)}^{\mu
_1\ldots \mu _{p-2k-1}}$, where $\bar \epsilon _{(p)}^{\mu _1\ldots \mu
_{p-2k-1}}=\Box \epsilon _{(p)}^{\mu _1\ldots \mu _{p-2k-1}}$, so the $\chi
_{(p)}$'s are purely gauge. Then, it is natural to take the gauge conditions 
$\chi _{(p)}^{\mu _1\ldots \mu _{p-2k-1}}=0$ in the gauge-fixing process.
They are enforced via the gauge-fixing fermion 
\begin{equation}
\label{64}\psi _{(p)}=\sum_{k=0}^{a_p}\int d^Dx\,\bar \eta _{(p)\mu _1\ldots
\mu _{p-2k-1}}\chi _{(p)}^{\mu _1\ldots \mu _{p-2k-1}}, 
\end{equation}
from which we derive the gauge-fixed action 
\begin{eqnarray}\label{65}
& &S_{\psi _{(p)}}^{(p)}=S_{0_p}^L+\sum_{k=0}^{a_p}\int d^Dx\left( \bar \eta
_{(p)\mu _1\ldots \mu _{p-2k-1}}\Box \eta _{(p)}^{\mu _1\ldots \mu
_{p-2k-1}}+\right. \nonumber \\  
& &\left. B_{(p)\mu _1\ldots \mu _{p-2k-1}}\chi _{(p)}^{\mu _1\ldots
\mu _{p-2k-1}}\right) .
\end{eqnarray}
It is easy to check that the gauge-fixed action (\ref{65}) possesses no
residual gauge invariances. The formula (\ref{65}) is the final output of
our irreducible procedure for abelian free $p$-form gauge fields. It
establishes that one can consistently quantize free abelian $p$-forms
without introducing either ghosts of ghosts or their antifields. At the same
time, we remark that our approach outlines some good gauge conditions with a
direct physical content. These conditions appear in the reducible procedure
from the necessity of implementing some irreducible Lorentz type gauge
conditions.

As can be seen from (\ref{51}), our procedure activates some of the
ineffective gauge transformations implied within the reducible treatment.
Then, it appears legitimate the question whether we can make effective all
the ineffective gauge transformations. The answer is however negative. We
show this in the simple case of three-form gauge fields, the general proof
following a similar line. The gauge invariances of abelian three-forms, $%
\delta _\epsilon A_{(3)}^{\mu \nu \rho }=\partial _{}^{\left[ \mu \right.
}\epsilon _{(3)}^{\left. \nu \rho \right] }$, become ineffective if one
takes $\epsilon _{(3)}^{\nu \rho }=\partial _{}^{\left[ \nu \right.
}\epsilon _{(3)}^{\left. \rho \right] }$ and further $\epsilon _{(3)}^\rho
=\partial ^\rho \epsilon _{(3)}$. Accordingly our method, only the gauge
transformations with the parameters $\epsilon _{(3)}^\rho =\partial ^\rho
\epsilon _{(3)}$ become effective via $\delta _\epsilon A_{(3)}^\mu
=\partial ^\mu \epsilon _{(3)}+2\partial _\nu \epsilon _{(3)}^{\nu \mu }$.
In order to make also effective the transformations with the gauge
parameters $\epsilon _{(3)}^{\nu \rho }=\partial _{}^{\left[ \nu \right.
}\epsilon _{(3)}^{\left. \rho \right] }$, it is necessary to add the new
fields $\left( A_{(3)}^{\mu \nu },A_{(3)}\right) $ with the gauge variations 
$\delta _\epsilon A_{(3)}^{\mu \nu }=\partial _{}^{\left[ \mu \right.
}\epsilon _{(3)}^{\left. \nu \right] }$, respectively, $\delta _\epsilon
A_{(3)}=\partial _\nu \epsilon _{(3)}^\nu $. The gauge transformations of
the theory with three-, two-, one-, and zero-form gauge fields are
irreducible. However, this new irreducible system does not describe the same
physical observables like those of free abelian three-form gauge fields.
This can be seen by writing down the equations associated with the physical
observables of the irreducible theory, namely, $3\partial ^\mu \frac{\delta F%
}{\delta A_{(3)}^{\mu \nu \rho }}+\partial _{\left[ \nu \right. }\frac{%
\delta F}{\delta A_{(3)}^{\left. \rho \right] }}=0$, $2\partial ^\mu \frac{%
\delta F}{\delta A_{(3)}^{\mu \nu }}+\partial _\nu \frac{\delta F}{\delta
A_{(3)}}=0$, $\partial ^\mu \frac{\delta F}{\delta A_{(3)}^\mu }=0$. From
the first and third equations, we infer $\partial ^\mu \frac{\delta F}{%
\delta A_{(3)}^{\mu \nu \rho }}=0$, $\frac{\delta F}{\delta A_{(3)}^\mu }=0$%
, while the second relation implies $\frac{\delta F}{\delta A_{(3)}}=0$, and 
$\partial ^\mu \frac{\delta F}{\delta A_{(3)}^{\mu \nu }}=0$. The last
equation does not lead to $\frac{\delta F}{\delta A_{(3)}^{\mu \nu }}=0$, so
the observables of the irreducible theory do not coincide with those of the
reducible one. We can remove this deficiency by adding a new gauge parameter 
$\epsilon _{(3)}^{\mu \nu \rho }$, and taking the gauge transformations of $%
A_{(3)}^{\mu \nu }$ under the form $\delta _\epsilon A_{(3)}^{\mu \nu
}=\partial _{}^{\left[ \mu \right. }\epsilon _{(3)}^{\left. \nu \right]
}+3\partial _\rho \epsilon _{(3)}^{\rho \mu \nu }$. In this situation, the
above equations fulfilled by the observables must be supplemented with $%
\partial _{\left[ \rho \right. }\frac{\delta F}{\delta A_{(3)}^{\left. \mu
\nu \right] }}=0$. Applying $\partial ^\rho $ on the last relation, and
using $\partial ^\mu \frac{\delta F}{\delta A_{(3)}^{\mu \nu }}=0$, it
follows $\frac{\delta F}{\delta A_{(3)}^{\mu \nu }}=0$, as required. Hence,
the supplementary gauge parameters $\epsilon _{(3)}^{\mu \nu \rho }$ helps
us at recovering the equivalence between the original and the new model at
the level of physical observables. A new problem appears now, namely, the
acyclicity of the Koszul-Tate operator. Indeed, $\delta $ is no longer
acyclic. On behalf of the definitions of $\delta $ acting on the antifields,
we find at the resolution degree equal with two the $\delta $-closed
quantities $\partial ^\lambda \partial _{\left[ \lambda \right. }\eta
_{(3)\left. \mu \nu \rho \right] }^{*}$, where $\eta _{(3)\mu \nu \rho }^{*}$
denote the antifields corresponding to the gauge parameters $\epsilon
_{(3)}^{\mu \nu \rho }$. In change, these quantities are not $\delta $%
-exact. This signalizes that the part of the gauge transformations for $%
A_{(3)}^{\mu \nu }$ involving the parameters $\epsilon _{(3)}^{\mu \nu \rho
} $ is vanishing under the change $\epsilon _{(3)}^{\mu \nu \rho }=\partial
_\lambda \partial ^{\left[ \lambda \right. }\theta _{(3)}^{\left. \mu \nu
\rho \right] }$ with arbitrary non-vanishing $\theta _{(3)}^{\mu \nu \rho }$%
's, which means that some reducibility is present. Thus, if we try to make
physically equivalent the two theories, then we lose the irreducibility,
and, conversely, if we ensure the irreducibility, the two systems are no
longer physically equivalent. In consequence, one cannot make effective all
the ineffective gauge transformations, and, at the same time, enforce the
irreducibility. This argues the introduction only of the fields $%
A_{(p)}^{\mu _1\ldots \mu _{p-2k}}$ with the gauge transformations (\ref{51}%
).

With the above analysis at hand, we are ready to investigate the irreducible
BRST quantization of gauge theories with abelian $p$-form gauge fields
involving interacting terms that are gauge invariant under (\ref{2}). These
terms may increase the derivative order of the field equations because the
Lagrangian density contains only the field strengths and their derivatives.%
\footnote{%
In certain dimensions, one can add topologically interactions. Such
interactions are not strictly gauge invariant, but only invariant up to some
surface terms.} For definiteness, we begin with a gauge theory described by
the action 
\begin{equation}
\label{66}S_0^L\left[ \left( A_{(p_a)}^{\mu _1\ldots \mu _{p_a}}\right)
\right] =\sum\limits_aS_{0_{p_a}}^L\left[ A_{(p_a)}^{\mu _1\ldots \mu
_{p_a}}\right] +S_I^L\left[ \left( A_{(p_a)}^{\mu _1\ldots \mu
_{p_a}}\right) \right] , 
\end{equation}
where $a=1,\ldots ,n$ and $p_a\geq 1$. Here, $A_{(p_a)}^{\mu _1\ldots \mu
_{p_a}}$ represents an abelian $p_a$-form, $S_{0_{p_a}}^L\left[
A_{(p_a)}^{\mu _1\ldots \mu _{p_a}}\right] $ is of the type (\ref{1}), and $%
S_I^L$ involves all the consistent interaction terms invariant under the
gauge transformations of the type (\ref{2}) for every $p_a$. The theory
described by (\ref{66}) can be quantized in an irreducible manner
accordingly the approach to free $p$-forms discussed earlier. In this light,
for every $p_a>1$ (abelian one-forms are irreducible) we add the fields $%
A_{(p_a)}^{\mu _1\ldots \mu _{p_a-2k}}$, $k=1,\ldots ,b_{p_a}$, the gauge
parameters $\epsilon _{(p_a)}^{\mu _1\ldots \mu _{p_a-2k-1}}$, $k=1,\ldots
,a_{p_a}$, and require the gauge transformations of the type (\ref{51}).
Although we investigate an interacting theory, the analysis goes almost
identically with the free case for every sort of original fields because the
reducibility functions of the interacting theory contain diagonal blocks.
The only difference resides in the action of the Koszul-Tate operator on the
initial fields, as the interaction terms may add some new terms to the free
equations of motion. This does not affect the analysis from the free case,
as the new equations of motion satisfy the same Noether identities like in
the absence of interaction. Hence, the proof of the physical equivalence (at
the level of observables) between the reducible and irreducible theories,
respectively, the acyclicity of the irreducible Koszul-Tate operator remains
unchanged. With these considerations at hand, it is simply to see that the
gauge-fixed action reads 
\begin{equation}
\label{67}S_\psi =S_0^L\left[ \left( A_{(p_a)}^{\mu _1\ldots \mu
_{p_a}}\right) \right] +\sum\limits_aS_{\psi _{p_a}}^{(p_a)}. 
\end{equation}
In (\ref{67}) $S_{\psi _{p_a}}^{(p_a)}$ is expressed by the second and third
terms in the right-hand side of (\ref{65}), with $p\rightarrow p_a$. The
gauge-fixing fermion on account of which we reach (\ref{67}) is given by $%
\psi =\sum\limits_a\psi _{p_a}$, with $\psi _{p_a}$ like in (\ref{64}). The
derivation of the gauge-fixed action (\ref{67}) completes our treatment. At
this point, we mention that our irreducible Lagrangian investigation of
interacting $p$-form gauge theories is advantageous as compared with a
Hamiltonian analysis because $S_I^L$ may contain higher-order derivative
terms which overwhelm both the canonical approach and the construction of
some irreducible first-class constraints.

To conclude with, in this paper we proved that gauge theories with abelian $%
p $-form gauge fields can be quantized along an irreducible antifield BRST
fashion. The cornerstone of our approach is given by the construction of an
irreducible gauge theory in a way that makes legitimate the replacement of
the reducible antifield-BRST quantization with the irreducible one. The
acyclicity of the irreducible Koszul-Tate operator was explicitly
emphasized, and also the equivalence between the irreducible and reducible
theories at the level of physical observables was completely elucidated. At
the same time, our formalism leads to some gauge conditions allowing a
meaningful physical interpretation. Our results will be used in a next paper
at the investigation of the deformation of the master equation \cite{19} for
gauge theories with abelian $p$-forms, and also at solving appropriately
some cohomological aspects linked with such theories.

\end{document}